\documentclass[runningheads]{llncs}

\usepackage{graphicx}
\usepackage{booktabs}
\usepackage{multirow}
\begin{document}
\title{Memory efficient brain tumor segmentation using an autoencoder-regularized U-Net}
\titlerunning{Memory efficient brain tumor segmentation}
\author{Markus Frey\inst{1} \and %\orcidID{0000-1111-2222-3333} \and
Matthias Nau\inst{1}} %\orcidID{1111-2222-3333-4444} \and

\institute{\textsuperscript{1}Kavli Institute for Systems Neuroscience, NTNU, Trondheim, Norway}
\maketitle % typeset the header of the contribution

% abstract
\begin{abstract}
Early diagnosis and accurate segmentation of brain tumors are imperative for successful treatment. Unfortunately, manual segmentation is time consuming, costly and despite extensive human expertise often inaccurate. Here, we present an MRI-based tumor segmentation framework using an autoencoder-regularized 3D-convolutional neural network. We trained the model on manually segmented structural T1, T1ce, T2, and Flair MRI images of 335 patients with tumors of variable severity, size and location. We then tested the model using independent data of 125 patients and successfully segmented brain tumors into three subregions: the tumor core (TC), the enhancing tumor (ET) and the whole tumor (WT). We also explored several data augmentations and preprocessing steps to improve segmentation performance. Importantly, our model was implemented on a single NVIDIA GTX1060 graphics unit and hence optimizes tumor segmentation for widely affordable hardware. In sum, we present a memory-efficient and affordable solution to tumor segmentation to support the accurate diagnostics of oncological brain pathologies.
\keywords{Brain tumor \and U-Net \and Autoencoder.}
\end{abstract}

% introduction
\section{Introduction}
An estimated 17.760 people will die from a primary brain tumor this year in the US alone \cite{ref_siegel}. Another 23.820 will be diagnosed with having one \cite{ref_siegel}. The earlier and the more accurate this diagnosis will be, the higher the patients chances are for successful treatment. In cases of doubt, patients typically undergo a brain scan either using computed tomography (CT) or magnetic resonance imaging (MRI). Both techniques acquire a 3D image of the brain, which then serves as the basis for medical examination. To understand the severity of the disease and to plan potential treatments, a critical challenge is identifying the tumor, but also to estimate its spread and growth by segmenting the affected tissue. This process still often relies on careful manual assessment by trained medical staff.\\

In recent years, a growing number of algorithmic solutions were proposed to aid and accelerate this process \cite{ref_brats1,ref_myro,ref_isensee}. Most of these automatic segmentation methods build on convolutional neural networks (CNNs) trained on manual brain segmentations of a large cohort of patients. Given enough training data, they learn to generalize across patients and allow to identify the tumor and its spread in new, previously unseen brains. However, there are at least two challenges associated with CNN's. First, they tend to overfit to the training data, making it necessary to either have large data sets to begin with, or to use a variety of data augmentations to make them generalize more robustly. Second, many current CNN implementations require powerful computational resources to be used within a reasonable time.\\

To solve such challenges and to promote the further development of automatic segmentation methods, the brain tumor segmentation challenge (BraTS) \cite{ref_brats1,ref_brats2,ref_brats3,ref_brats4,ref_brats5} provides large data sets of manually segmented brains for users to test new implementations. Here, we used this data to implement a convolutional autoencoder regularized U-net for brain tumor segmentation inspired by last year's BraTS challenge winning contribution \cite{ref_myro}. As model input, we used structural (T1) images, T1-weighted contrast-enhanced (T1ce) images, T2-weighted images and fluid-attenuated inversion recovery (Flair) MRI images of 335 patients with tumors of variable severity, size and location. As training labels, we used the corresponding manual segmentations.\\ 

The model training comprised three parts (Fig. 1). First, in an encoding stage, the model learned a low-dimensional representation of the input. Second, the variational autoencoder (VAE) stage reconstructed the input image from this low-dimensional latent space. Third, a U-Net part created the actual segmentations \cite{ref_unet}. In this model architecture, the VAE part is supposed to act as a strong regularizer on all model weights \cite{ref_myro} and therefore to prevent overfitting on the training data. The resulting segmentation images were compared to the manual segmentation labels. This process was repeated until the optimal model weights were found. These optimal parameters were then tested on new validation data of 125 patients, localizing and segmenting each brain tumor into three tissue categories: whole tumor, enhancing tumor and tumor core.\\

Importantly, all of these steps were conducted on a single NVIDIA GTX1060 graphics unit while using data exclusively from the BraTS challenge 2019. In addition, we explored various model parameters, data augmentations and preprocessing steps to improve model performance. Therefore, we address above-introduced challenges by presenting a memory-efficient and widely-affordable solution to brain tumor segmentation in line with the aims of the GreenAI initiative \cite{ref_greenai}.

% methods
\section{Methods}

\subsection{Model architecture}

\begin{figure}
\includegraphics[width=1\textwidth]{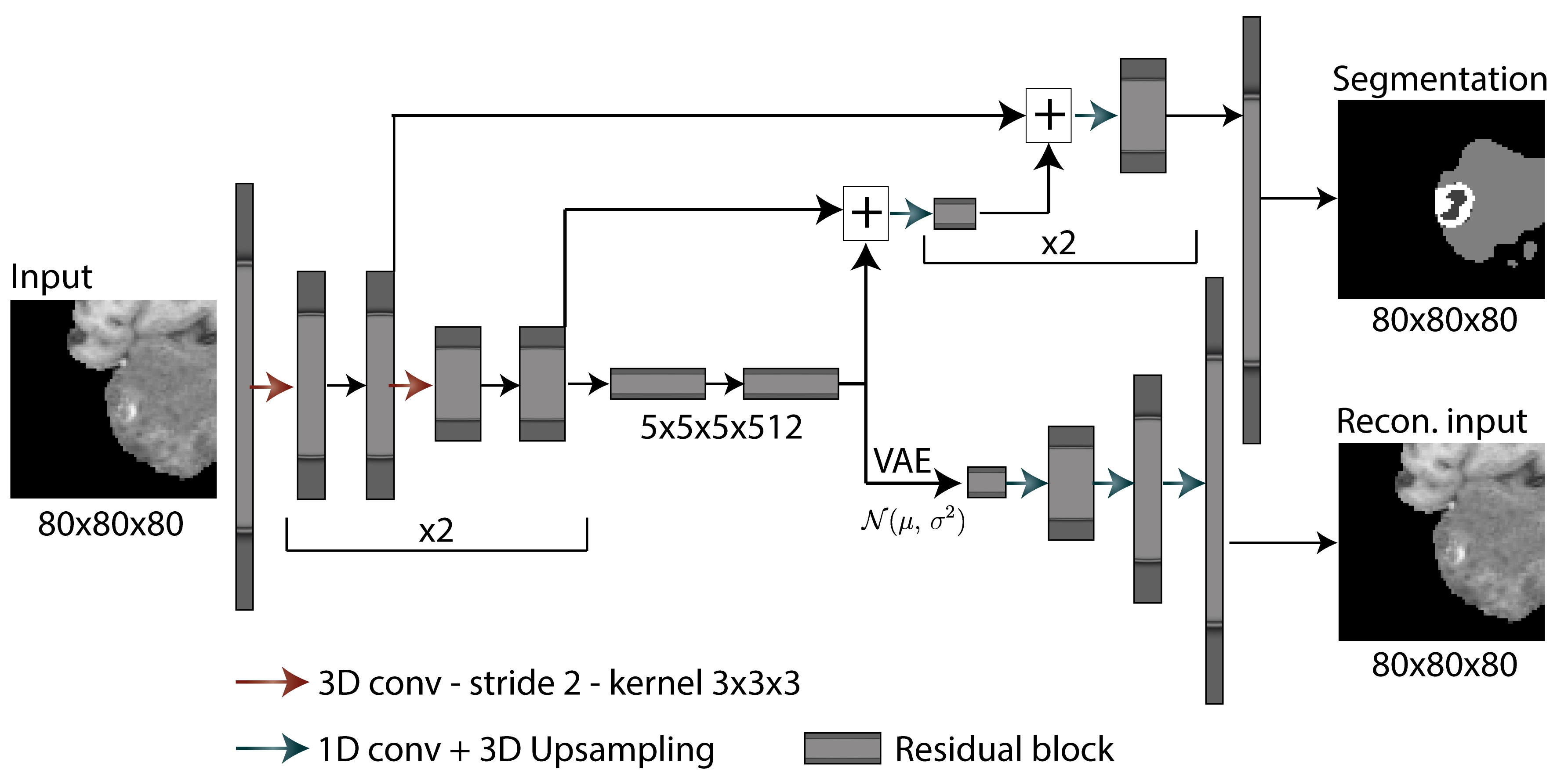}
	\caption{Model architecture of our memory-efficient autoencoder-regularized U-Net. As input to the model we used patches of size 80x80x80 and stacked the MRI modalities in the channel dimension (n=4). We used 3D convolutions with a kernel size of 3x3x3 throughout. We used residual blocks \cite{ref_resnet}, using 3D convolutions with LeakyReLU activations, interspersed with Group Normalization \cite{ref_groupnorm}. For upsampling to the original image size, we used 3D bilinear upsampling and 3D convolutions with a kernel size of 1 for both the autoencoder and the segmentation part.}
	\label{fig_1}
\end{figure}

As mentioned above, our model is inspired by earlier work \cite{ref_myro}, but was adapted as described in the following (see also Figure \ref{fig_1}). We adjusted the model architecture to incorporate a patch-wise segmentation of the input image, as the full input with a resolution of 240x240x155 voxel as used in the original model is too big to fit most commercially available graphics cards (GPU). This is true even with a batch size of 1. We therefore used 3D blocks of size 80x80x80 and adjusted the number of filters to make full use of the GPU memory available, leading to 32 filters in the first layer with a ratio of 2 between subsequent layers. We also replaced the rectified linear unit (ReLU) activation functions with LeakyReLU \cite{ref_lrelu} as we observed an improvement in performance in a simplified version of our model.\\

Notably, we tested various other factors, which did not lead to an improvement in model performance, but are nevertheless included here as null-report. These included a) changing the downsampling in the convolutional layers from strides to average or max pooling, b) adjusting the ratio in the number of filters between layers (including testing non-integer steps), c) varying the number of units in the bottleneck layer, d) increasing the number of down-sampling and subsequent up-sampling steps and e) replacing the original group norm by batch norm. Due to our self-imposed computational constraints, we could not systematically test all these adjustments and possible interactions using the full model. Instead, we tested these parameters in a simplified model with only 8 filters at the input stage.\\

The overall model architecture follows a similar structure as a U-Net \cite{ref_unet}, with an additional variational autoencoder module \cite{ref_kingma} to regularize the segmentation of the tumor masks. As loss functions for the autoencoder we used the mean-squared error between the reconstructed and real input image and the Kullback-Leibler loss to ensure a normal distribution in the latent space. The weights for both losses were down-weighted by a factor of 0.1. The (soft Dice) segmentation loss was averaged across all voxels belonging to the whole tumor (WT), enhancing tumor (ET) and tumor core (TC).\\

% Add a figure:

\subsection{Optimization}
For training the model we used an adjusted version of the Dice loss in \cite{ref_myro}:

\begin{equation}
L_{Dice} = 1 - \Big(2 * \frac{\sum (y_{true} * y_{pred}) + s} {(\sum y_{true}^2 + \sum y_{pred}^2) + s}\Big)
\end{equation}

with $y_{true}$ being the real 3D mask and $y_{pred}$ being the corresponding 3D prediction. This version of the Dice loss ensured that the loss estimate lies within the interval [0,1]. The smoothness term $s$ ensured that the model is allowed to predict 0 tumor voxels without incurring a high loss in its overall estimate. In line with \cite{ref_cahall} we decided to use $s = 100$. \\

\newpage
The autoencoder part of our model consisted of two loss terms. As a reconstruction loss we used the mean-squared error between the reconstructed and the real input image:

\begin{equation}
L_{L2} = || y_{true} - y_{pred} ||^2_2 
\end{equation}

In addition, we used a Kullback-Leibler loss to ensure a normal distribution in our bottleneck layer, with N being the number of voxels in the input: 

\begin{equation}
L_{KL} = \frac{1}{N}\sum \mu^2 + \sigma^2 - \log \sigma^2 - 1
\end{equation}

with $\mu$ and $\sigma^2$ the mean and variance of the estimated distribution. In line with \cite{ref_myro} we weighted the autoencoder losses by 0.1, resulting in an overall loss according to:

\begin{equation}
L = 0.1 * L_{L2} + 0.1 * L_{KL} + 0.33 * L_{Dice_{wt}} + 0.33 * L_{Dice_{tc}} + 0.33 * L_{Dice_{et}}
\label{eq_loss}
\end{equation}

We tested different weighting for the tumor subregions, but did not observe a clear change in model performance using the smaller test model. We therefore used the average of the three regions. 

\newpage
For training the model, we used the Adam optimizer \cite{ref_adam}, starting out with a learning rate of 1e-4 and decreasing it according to

\begin{equation}
\alpha = \alpha_0 * (1 - \frac{e}{N_e})^{0.9}
\end{equation}

with $e$ the epoch and $N_e$ the number of total epochs (n=50). We evaluated 2101 samples in each epoch, stopping early when the validation loss did not decrease further for 2 subsequent epochs. 

\subsection{Data augmentation}
\begin{figure}[ht]
\includegraphics[width=1\textwidth]{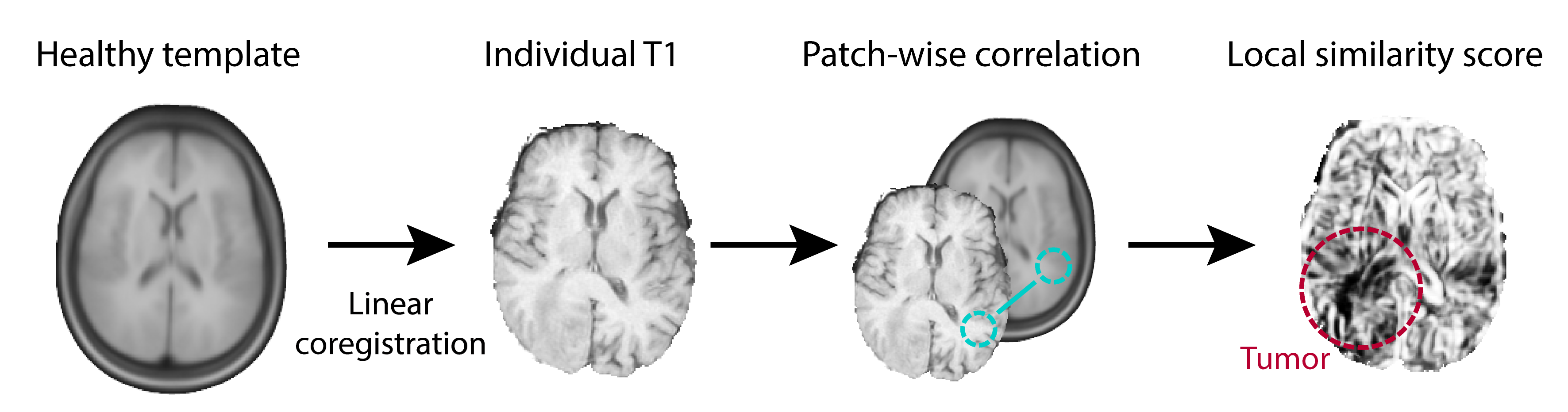}
	\caption{Local similarity score. To aid model performance, we computed a local similarity score image, which served as additional input. We linearly co-registered a healthy template brain \cite{ref_evans} to each participant's T1, and computed a patch-wise Pearson correlation between the two. Patch-size was 7mm. The correlation between healthy and pathological brain drops in tumor regions.}\label{fig_2}
\end{figure}

In line with \cite{ref_myro} we used a random scaling between 0.9 and 1.1 on each image patch, and applied random axis mirror flip for all 3 axes with a probability of 0.5. We experimented with additional augmentations. In particular, we computed a voxel-wise similarity score for each participant's T1 comparing it to a healthy template brain. We co-registered an average template of 305 participants without any tumors \cite{ref_evans} to each patient's T1 using translations, rotations and scaling and calculated a patch-wise Pearson's correlation with a searchlight-sphere size of 7mm. The resulting correlation images were normalized and concatenated with the 4 MRI modalities as an additional channel in the input (Figure \ref{fig_2}). However, in our hands, this procedure did not further improve model performance. Future work could test different across-image similarity measures.\\

% Add a figure:

A shortcoming of using discretized patch-input is the lack of information about the anatomical symmetry of the tested brain images. Strong asymmetry in MRI images can indicate the presence of a tumor, which is in most cases limited to one hemisphere. The other hemisphere should hence rather approximate how the healthy brain once looked like. Therefore, for each patch we also provided the mirrored patch from the opposite hemisphere as an additional input. This mirroring of image patches was only done on the x-axis of the MRI images. Looking forward, we believe this approach has the potential to benefit the model performance if measures other than Pearson's correlation are explored and mirror symmetry is factored in.\\

We used test time augmentation to make our segmentation results more robust, for this we mirrored the input on all three axes and flipped the corresponding prediction to match the original mask orientation. This gave us 16 model estimates (2 * 2 * 2 * 2), which we averaged and thresholded to obtain our segmentation masks. We decided to directly optimize for tumor regions instead of the intra-tumoral regions as this resulted in better estimates during training of how well our model will perform on the BraTS 2019 competition benchmark. We optimized the values at which we thresholded our masks on the training data and used 0.55, 0.5 and 0.4 for whole tumor, tumor core and enhanced tumor respectively. \\

\section{Results}
Here, we present an autoencoder regularized U-net for brain tumor segmentation. The model was trained on the BraTS 2019 training data, which consisted of 335 patients separated into high-grade glioma and low-grade glioma cases. Initial shape of the input data was 240x240x155, with multi-label segmentation masks of the same size, indicating NCR \& NET (label 1), edema (label 2), and enhancing tumor (label 4). We created an average tumor template from all segmentation masks to locate the most prominent tumor regions via visual inspection. Based on that, we created our initial slice resulting in image dimensions of 160x190x140. We then used a sliding window approach to create patches of size 80x80x80, feeding these patches through the model while using a sampling procedure that increased the likelihood of sampling patches with a corresponding tumor (positive samples).\\

\begin{table}[ht]
\centering
\scriptsize
    \begin{tabular}{c@{\qquad}ccc@{\qquad}ccc@{\qquad}ccc@{\qquad}ccc}
      \toprule
      \multicolumn{3}{l}{Dice score} & \multicolumn{3}{l}{Sensitivity} & \multicolumn{3}{l}{Specificity} & \multicolumn{3}{l}{Hausdorff95}\\
      \cmidrule{2-13}
      & ET & WT & TC & ET & WT & TC & ET & WT & TC & ET & WT & TC \\
      \midrule
      Mean & 0.787 & 0.896 & 0.800 & 0.782 & 0.907 & 0.787 & 0.998 & 0.994 & 0.997 & 6.005 & 8.171 & 8.241 \\
      Std & 0.252 & 0.085 & 0.215 & 0.271 & 0.088 & 0.246 & 0.003 & 0.008 & 0.004 & 14.55 & 15.37 & 11.53 \\
      Median & 0.870 & 0.922 & 0.896 & 0.884 & 0.934 & 0.895 & 0.999 & 0.997 & 0.999 & 2.000 & 3.162 & 3.605 \\
      \bottomrule
    \end{tabular}
    \hspace{10pt}
    \caption{Validation results. ET: enhancing tumor, WT: whole tumor, TC: tumor core.}
    \label{table_1}
\end{table}
\vspace{-5mm}

\begin{figure}[ht]
\includegraphics[width=0.9\textwidth]{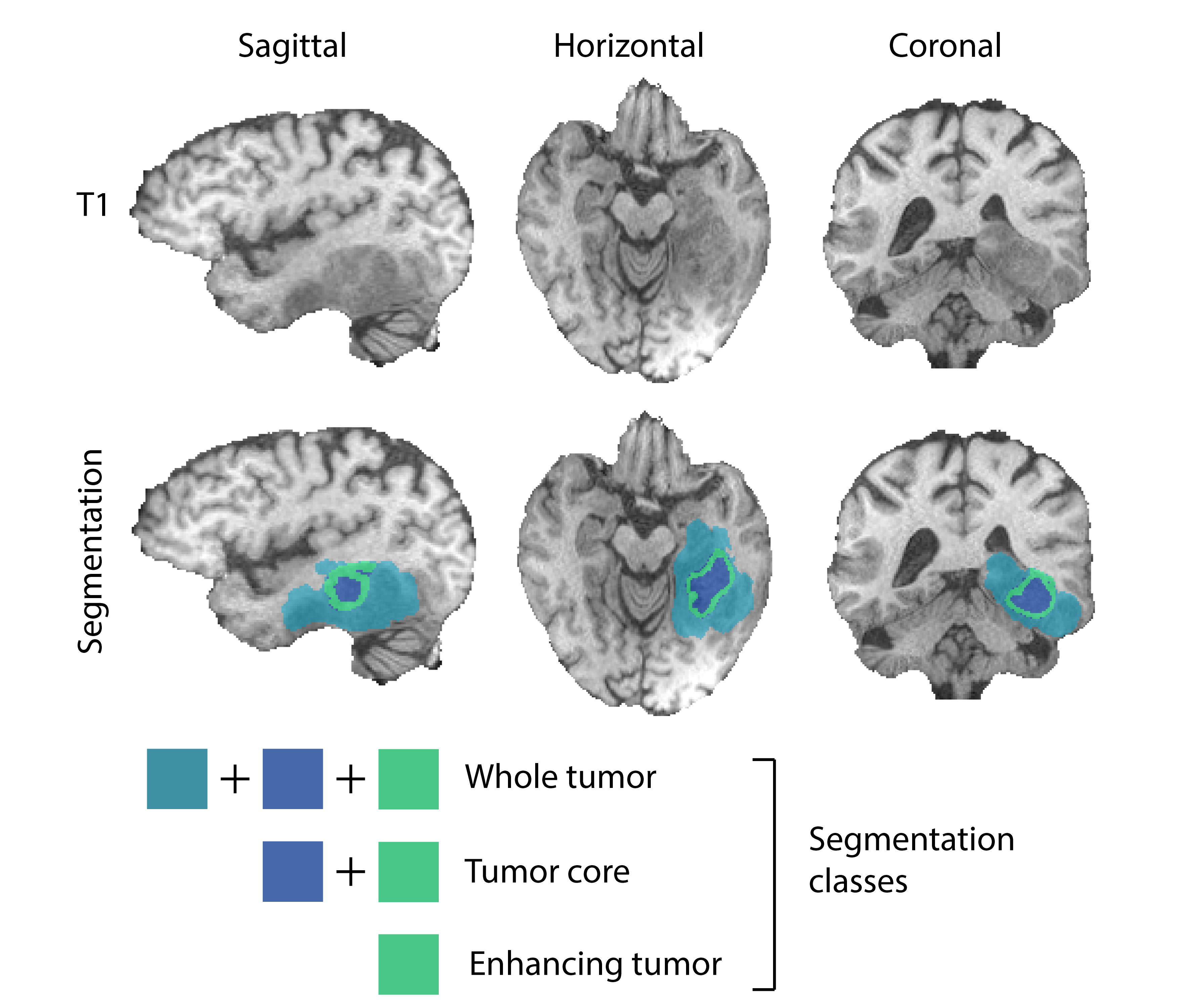}
	\caption{Tumor segmentations. Validation data shown for one exemplary patient. We depict the T1 scan (upper panel) as well as the segmentation output of our model overlaid on the respective T1 scan (bottom panel) for sagittal, horizontal and coronal slices. Segmentations were color-coded.}\label{fig_3}
\end{figure}

\begin{figure}[ht]
\includegraphics[width=0.9\textwidth]{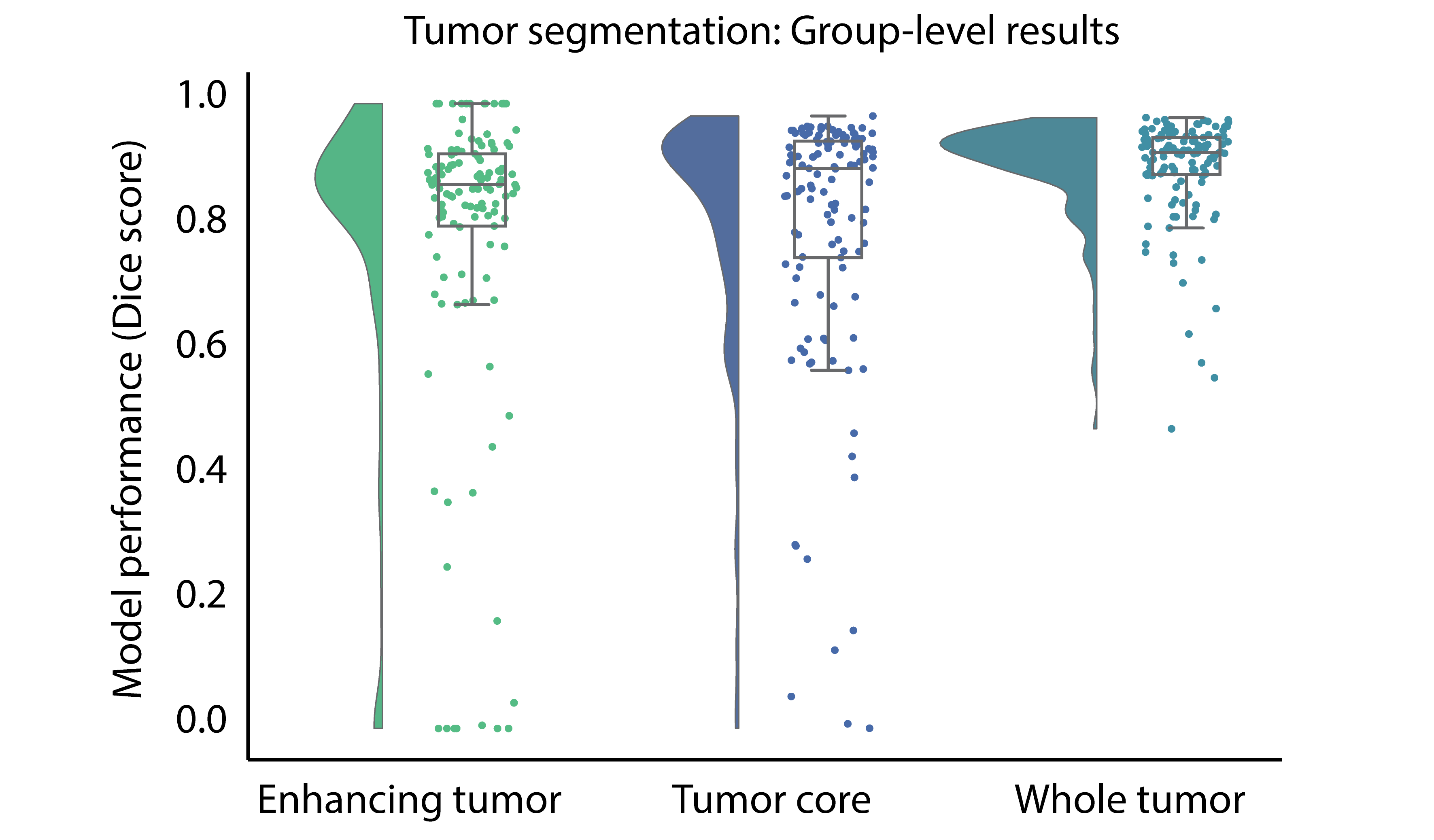}
	\caption{Group-level segmentation performance (Dice score) for enhancing tumor (green, left), tumor core (blue, middle) and whole tumor (petrol, right) for the validation data set. We plot single-patient data overlaid on group-level whisker-boxplots (center, median; box, 25th to 75th percentiles; whiskers, 1.5 × interquartile range) as well as the smoothed data distribution.}\label{fig_4}
\end{figure}

We used an ensemble of two separately trained models to segment the MRI images of validation and testing set into different tumor tissue types. This allowed us to test the model on previously unseen data (Table \ref{table_1}, Figure \ref{fig_3}, \ref{fig_4}, team-name: CYHSM). The mean Dice scores of our model on the validation dataset (n=125) are 0.787 for enhanced tumor, 0.896 for whole tumor and 0.800 for tumor core. \\

% figure caption:

In Figure \ref{fig_3}, we show the model segmentations for one exemplary patient from the validation set overlayed on the patient's T1 scan. For this patient we obtained Dice scores of 0.923 for whole tumor, 0.944 for tumor core and 0.869 for enhancing tumor from the online evaluation platform: https://ipp.cbica.upenn.edu. The distribution of Dice scores across patients can be seen in Figure \ref{fig_4} \cite{ref_raincloud}. The ensemble model performed well on most patients ($\sim$ 0.9 median Dice score), but failed completely in a few.\\

\newpage
\begin{figure}[ht]
\centering
\includegraphics[width=0.8\textwidth]{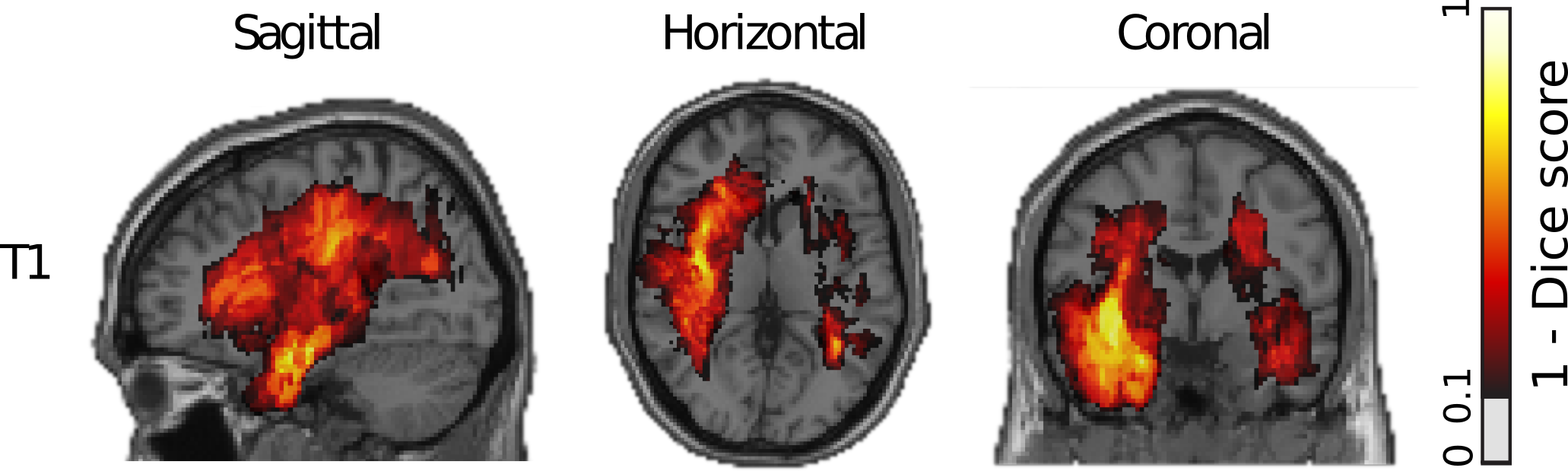}
	\caption{Model error (1-Dice Score) overlaid on structural T1-template in MNI-space. Hot colors indicate high errors. The model performed well in superficial gray matter, but failed in deeper structures, especially in white matter tracts in the temporal lobe.}\label{fig_5}
\end{figure}

To examine why the model performed poorly in some few patients, we examined the model error pattern as a function brain location. We calculated the average voxel-wise Dice score for the whole tumor for all 125 validation subsets and registered them to the Colin27-MNI-template (Figure \ref{fig_5}). We found that our model performed well in superficial gray matter (average Dice-score $>$0.9), but failed to segment the tumors accurately in white matter, predominantly in deeper structures in the temporal lobes. Moreover, our model segmented the whole tumor most accurately, but struggled to differentiate the enhancing tumor from the tumor core. It especially misclassified low-grade glioma cases in which no enhancing tumor was present (Dice score of 0). 

\section{Discussion}
Tumor segmentation still often relies on manual segmentation by trained medical staff. Here, we present a fast, automated and accurate solution to this problem. Our segmentations can be used to inform physicians and aid the diagnostic process. We successfully segmented various brain tumors into three tissue types: whole tumor, enhancing tumor and tumor core in 125 patients provided by the BraTS challenge \cite{ref_brats1}. Importantly, our model was implemented and optimized on a single GTX1060 graphics unit with 6GB memory. To meet these low graphics memory demands, we split the input images into multiple 3D patches. The model iterated through these patches and converged on the most likely brain segmentation given all iterations in the end. We hence present a memory efficient and widely affordable solution to brain segmentation. Naturally, one limitation of this low-cost approach is that the model is still relatively slow. Naturally, more computational resources would alleviate this problem. In addition, more graphics memory would allow to upscale the input patch size further, in turn likely also benefiting the model performance greatly.\\

In addition, we implemented the model using data provided for this year's BraTS 2019 challenge alone. No other data was used. Earlier work including previous BraTS challenges showed that incorporating additional data, hence increasing the training data set, greatly reduces overfit and improves model performance drastically \cite{ref_isensee}. Here, we aimed at optimizing brain tumor segmentation explicitly in the light of these common computational and data resource constraints. One interesting observation was that the model performed well on most patients (\ref{fig_3}), but failed completely in a few. The reasons for this remain unclear and need to be explored in the future.\\

Taken together, our results demonstrate the wide-ranging applicability of U-Nets to improve tissue segmentation and medical diagnostics. We show that dedicated memory efficient model architectures can overcome computational and data resource limitations and that fast and efficient brain tumor segmentation can be achieved on widely-affordable hardware.\\

\noindent \textbf{\large Acknowledgements}\\

\noindent We are grateful to Christian F. Doeller and the Kavli Institute for Systems Neuroscience for supporting this work.

% ---- Bibliography ----

\end{document}